\documentclass[conference]{IEEEtran}

\usepackage[utf8]{inputenc}
\usepackage[T1]{fontenc}
\usepackage{amsmath,amssymb}
\usepackage{graphicx}
\usepackage{booktabs}
\usepackage{array}
\usepackage{xcolor}
\usepackage{url}
\usepackage{cite}
\usepackage{tikz}
\usetikzlibrary{arrows.meta, positioning, shapes.geometric, fit}

\graphicspath{{figures/}}

\title{Developing an Intelligent Job Recommendation System Using Semantic Retrieval and Explainable AI Techniques}

\author{
\IEEEauthorblockN{Hussein Ali Al Awad\IEEEauthorrefmark{1} \quad Dr. Khaled Fathi Omar\IEEEauthorrefmark{1}}
\IEEEauthorblockA{\IEEEauthorrefmark{1}Master of Web Science, Syrian Virtual University, Damascus, Syria\\
Hussein Ali Al Awad: \url{husseinite98@gmail.com}\\
Dr. Khaled Fathi Omar: \url{Kh.om.mail@gmail.com}}
}

\begin{document}

\maketitle

\begin{abstract}
Online recruitment platforms require recommendation methods that can retrieve relevant job opportunities from large and heterogeneous collections of postings. Keyword-based search is efficient and interpretable, but it can miss relevant postings when equivalent roles are expressed with different terms. This study presents a metadata-only job recommendation system that combines TF-IDF lexical matching, Sentence-BERT semantic retrieval, query-aware filtering, optional Cross-Encoder re-ranking, and explanation generation. The system uses structured fields, including job title, company name, location, seniority level, job function, employment type, and industry, without relying on full job descriptions or user interaction histories. Experiments on a cleaned LinkedIn job posting dataset containing 31,262 records show that the best hybrid configuration achieved Precision@10 of 0.8032 and nDCG@10 of 0.9496. Under the internal evaluation protocol, Cross-Encoder re-ranking improved Precision@10 from 0.7896 to 0.7948 and nDCG@10 from 0.9666 to 0.9739. These results indicate that lexical and semantic retrieval can be combined effectively for explainable job recommendation when only structured metadata is available.
\end{abstract}


\section{Introduction}

Online recruitment platforms have become important channels for matching job seekers with employment opportunities. These platforms commonly contain postings that differ by job title, company, location, seniority level, employment type, job function, and industry. As the number of postings increases, effective retrieval becomes necessary because users must identify relevant opportunities from large and heterogeneous collections.

Many recruitment search interfaces still rely on lexical matching. This approach is computationally efficient and transparent, and it remains useful for exact constraints such as location, seniority level, and employment type. However, lexical search can under-rank relevant postings when equivalent roles are expressed with different surface forms. For example, postings titled ``software engineer'', ``application developer'', and ``backend engineer'' may be relevant to a query for ``software developer'', even when exact term overlap is limited.

Dense semantic retrieval provides a complementary retrieval signal. Transformer-based language models, including BERT \cite{devlin2019bert} and Sentence-BERT \cite{reimers2019sentencebert}, can encode short texts into vector representations that support semantic similarity search. Such models are useful when job titles and metadata contain related meanings but different terms. Nevertheless, semantic similarity alone may be insufficient for job search because explicit constraints, including work mode, location, seniority, and employment type, should be preserved.

The gap addressed in this study is the design of an accurate and explainable job recommendation method for a restricted metadata-only setting. In this setting, full job descriptions, user profiles, click logs, application histories, and manually labeled relevance judgments are unavailable. The proposed system addresses this gap by combining TF-IDF lexical matching, Sentence-BERT semantic candidate retrieval, weighted hybrid ranking, query-aware filtering, optional Cross-Encoder re-ranking, and metadata-based explanation generation.

The main contributions of this paper are as follows:
\begin{itemize}
    \item A metadata-only job recommendation framework that does not require full descriptions or historical user interactions.
    \item A hybrid retrieval model that combines TF-IDF lexical similarity and Sentence-BERT semantic embeddings.
    \item A semantic candidate generation strategy based on normalized dense vectors and nearest-neighbor retrieval.
    \item Query-aware filtering for employment type, seniority level, location, and remote, hybrid, or onsite work mode.
    \item An explainability layer that reports matched terms, applied filters, and metadata-based evidence.
    \item An empirical evaluation using Precision@10 and nDCG@10 on a cleaned LinkedIn job posting dataset.
\end{itemize}

\section{Related Work}

Recommendation systems are commonly categorized as collaborative filtering, content-based filtering, or hybrid methods \cite{ricci2015recommender}. Collaborative filtering uses interaction histories such as ratings, clicks, or applications. Although this approach can be effective on mature platforms, it is limited in cold-start scenarios where new users or items have little historical data. Content-based methods use item attributes and are therefore more suitable when interaction data is unavailable.

Information retrieval methods provide a foundation for content-based recommendation. Classical lexical models such as TF-IDF are widely used because they preserve exact term evidence and can be implemented efficiently \cite{manning2008ir}. In job recommendation, exact terms remain important because many user queries contain explicit requirements, including locations, employment types, seniority levels, and work-mode constraints.

Transformer-based models have improved semantic representation and ranking. The transformer architecture introduced attention-based sequence modeling \cite{vaswani2017attention}, and BERT introduced bidirectional contextual representations for natural language processing \cite{devlin2019bert}. Sentence-BERT adapted transformer representations for efficient sentence-level similarity search \cite{reimers2019sentencebert}. Dense vector search can be scaled using nearest-neighbor indexing approaches, including GPU-oriented similarity search methods such as FAISS \cite{johnson2019billion}. Cross-Encoder ranking models jointly process a query and candidate document, often improving ranking quality at higher computational cost \cite{thakur2021beir,lin2021pretrained}. For this reason, they are typically applied to a limited set of top-ranked candidates rather than the full collection.

Explainability is also relevant to recommender systems, particularly in domains where recommendations may influence important decisions. Explanations may be generated through matched features, score decomposition, rules, or concise textual rationales \cite{zhang2020explainable}. In employment recommendation, metadata-based explanations are practical because visible fields such as job title, function, seniority, employment type, and location can be directly connected to user intent.

\section{Problem Statement}

This study addresses the problem of recommending relevant job postings using only structured metadata. The available fields include job title, company name, location, hiring status, date, seniority level, job function, employment type, and industry. Full job descriptions, user profiles, click logs, application histories, and human relevance labels are not assumed to be available.

This setting is challenging for several reasons. Job metadata is short, role titles are inconsistent across organizations, abbreviations are common, and user queries may combine semantic intent with strict constraints. A query such as ``remote junior data analyst London'' includes a target role, work mode, seniority level, and location. A useful recommender should capture semantic similarity for the role while preserving exact constraints when they are explicitly stated.

The main technical challenges are summarized as follows:
\begin{itemize}
    \item Lexical retrieval may miss semantically related postings with different wording.
    \item Semantic retrieval may return broadly related postings that violate explicit constraints.
    \item Metadata-only records provide less textual evidence than full job descriptions.
    \item Neural recommendation components require interpretable outputs to support user trust.
    \item Offline evaluation is difficult when human relevance labels are unavailable.
\end{itemize}

\section{Proposed Methodology}

The proposed methodology follows a pipeline-oriented design. Raw job postings are validated and cleaned, and each posting is converted into a composite metadata document. Two indexes are then created: a sparse TF-IDF matrix and a dense Sentence-BERT embedding matrix. During recommendation, the query is encoded using both representations. Semantic nearest-neighbor retrieval generates candidate postings, and lexical and semantic scores are normalized and combined through a weighted hybrid scoring function. Query-aware filters, duplicate suppression, company-level diversification, optional Cross-Encoder re-ranking, and explanation generation are then applied.

\subsection{Metadata Document Construction}

Each job posting is represented by a composite metadata document:
\begin{equation}
d_i = [t_i, t_i, c_i, l_i, s_i, f_i, e_i, r_i],
\end{equation}
where $t_i$ is the job title, $c_i$ is the company name, $l_i$ is the location, $s_i$ is the seniority level, $f_i$ is the job function, $e_i$ is the employment type, and $r_i$ is the industry. The title is repeated because it is usually the strongest visible signal in job search.

\subsection{Lexical Similarity}

The lexical component represents metadata documents and queries using TF-IDF. The TF-IDF weight of term $t$ in document $d$ is defined as \cite{manning2008ir}:
\begin{equation}
\operatorname{tfidf}(t,d) =
\operatorname{tf}(t,d)
\times
\log\left(\frac{N}{\operatorname{df}(t)+1}\right),
\end{equation}
where $N$ is the number of documents and $\operatorname{df}(t)$ is the number of documents containing term $t$. Lexical similarity is computed as a sparse dot product between the query vector and candidate document vectors.

\subsection{Semantic Similarity}

The semantic component uses the Sentence Transformer model \texttt{sentence-transformers/all-MiniLM-L6-v2}. Each document $d_i$ and query $q$ is encoded into normalized dense vectors $\mathbf{v}_{d_i}$ and $\mathbf{v}_q$. Semantic similarity is computed as:
\begin{equation}
s_{\mathrm{sem}}(q,d_i)=\mathbf{v}_q^\top \mathbf{v}_{d_i}.
\end{equation}
Because the embeddings are normalized, the inner product is equivalent to cosine similarity.

\subsection{Hybrid Ranking}

The lexical and semantic scores are min-max normalized before fusion:
\begin{equation}
\hat{s}(x)=\frac{s(x)-\min(s)}{\max(s)-\min(s)+\epsilon}.
\end{equation}
The hybrid score is then computed as:
\begin{equation}
s_{\mathrm{hybrid}}(q,d_i)=0.4\,\hat{s}_{\mathrm{sem}}(q,d_i)+0.6\,\hat{s}_{\mathrm{lex}}(q,d_i).
\label{eq:hybrid}
\end{equation}
The lexical component is assigned a larger weight because exact terms in titles, locations, work modes, and seniority levels are important in job search. The semantic component remains necessary because it expands retrieval beyond exact keyword overlap.

\subsection{Cross-Encoder Re-ranking}

When re-ranking is enabled, the top candidates are processed as query-document pairs using \texttt{cross-encoder/ms-marco-MiniLM-L-6-v2}. The final score is computed as:
\begin{equation}
s_{\mathrm{final}}(q,d_i)=
\alpha\,\hat{s}_{\mathrm{rerank}}(q,d_i)
+(1-\alpha)\,s_{\mathrm{hybrid}}(q,d_i)+b_i,
\label{eq:final}
\end{equation}
where $\alpha=0.7$ and $b_i$ is an optional metadata bonus assigned when a candidate shares strong metadata evidence with the seed job in the evaluation setting.

\subsection{Implementation Details}

The implementation used Python with \texttt{pandas}, \texttt{NumPy}, \texttt{scikit-learn}, \texttt{SciPy}, \texttt{joblib}, \texttt{sentence-transformers}, \texttt{PyTorch}, and \texttt{Streamlit}. Data preparation reads the LinkedIn Excel file, validates required columns, normalizes text, expands common abbreviations, removes noisy job titles, and saves the cleaned dataset as an artifact. Index training builds a TF-IDF matrix and a Sentence-BERT embedding matrix over the composite metadata documents. Runtime recommendation loads the stored artifacts, retrieves semantic candidates using \texttt{NearestNeighbors} with cosine distance, computes lexical and semantic scores, applies hybrid fusion, optionally performs Cross-Encoder re-ranking, removes duplicates, limits repeated companies, and returns explanation fields.

\section{System Architecture}

The architecture is organized as a layered retrieval and recommendation pipeline. Each layer has a specific function and passes structured outputs to the next layer. This modular design supports maintainability, reproducibility, and controlled experimentation.

\begin{table}[t]
\centering
\caption{Functional layers of the proposed job recommendation architecture.}
\label{tab:architecture}
\footnotesize
\begin{tabular}{@{}p{0.22\linewidth}p{0.42\linewidth}p{0.22\linewidth}@{}}
\toprule
\textbf{Layer} & \textbf{Main Function} & \textbf{Output} \\
\midrule
Data ingestion & Reads and validates structured job metadata & Validated records \\
Preprocessing & Normalizes text, expands abbreviations, and removes noisy titles & Cleaned metadata \\
Indexing & Builds TF-IDF vectors and Sentence-BERT embeddings & Sparse and dense indexes \\
Retrieval & Retrieves semantic candidates using vector similarity & Candidate list \\
Hybrid ranking & Combines lexical, semantic, metadata, and optional re-ranking scores & Ranked recommendations \\
Explainability & Reports matched terms, filters, and metadata evidence & Transparent output \\
\bottomrule
\end{tabular}
\end{table}

\begin{figure*}[t]
\centering
\resizebox{0.98\textwidth}{!}{%
\begin{tikzpicture}[
    node distance=0.72cm and 0.85cm,
    process/.style={rectangle, rounded corners=2pt, draw=blue!55!black, fill=blue!6, very thick, align=center, minimum width=2.8cm, minimum height=0.78cm, font=\scriptsize},
    store/.style={cylinder, shape border rotate=90, aspect=0.25, draw=teal!55!black, fill=teal!7, very thick, align=center, minimum width=2.65cm, minimum height=0.95cm, font=\scriptsize},
    output/.style={rectangle, rounded corners=2pt, draw=green!45!black, fill=green!8, very thick, align=center, minimum width=2.9cm, minimum height=0.78cm, font=\scriptsize},
    decision/.style={diamond, draw=orange!65!black, fill=orange!10, very thick, align=center, aspect=1.9, inner sep=1.2pt, font=\scriptsize},
    arrow/.style={-{Latex[length=2.2mm]}, very thick, draw=black!70},
    dashedarrow/.style={-{Latex[length=2.2mm]}, very thick, dashed, draw=black!55},
    group/.style={draw=black!35, rounded corners=4pt, dashed, inner sep=0.23cm}
]
\node[store] (raw) {Raw LinkedIn\\metadata};
\node[process, right=of raw] (prep) {Preprocessing\\cleaning and normalization};
\node[process, right=of prep] (doc) {Composite metadata\\document construction};
\node[process, above right=0.55cm and 0.9cm of doc] (tfidf) {TF-IDF\\indexing};
\node[process, below right=0.55cm and 0.9cm of doc] (sbert) {SBERT embedding\\generation};
\node[store, right=0.95cm of tfidf] (sparse) {Sparse\\TF-IDF matrix};
\node[store, right=0.95cm of sbert] (dense) {Dense embedding\\matrix};

\draw[arrow] (raw) -- (prep);
\draw[arrow] (prep) -- (doc);
\draw[arrow] (doc) -- (tfidf);
\draw[arrow] (doc) -- (sbert);
\draw[arrow] (tfidf) -- (sparse);
\draw[arrow] (sbert) -- (dense);

\node[process, below=2.0cm of prep] (query) {User query};
\node[process, right=of query] (qproc) {Query normalization\\and filter extraction};
\node[process, right=of qproc] (qenc) {Lexical and semantic\\query encoding};
\node[process, right=of qenc] (retrieval) {Semantic candidate\\retrieval};
\node[process, right=of retrieval] (ranking) {Hybrid ranking\\TF-IDF + SBERT};
\node[decision, right=of ranking] (rerank) {Cross-Encoder\\re-rank?};
\node[process, above right=0.1cm and 0.75cm of rerank] (ce) {Top-100 neural\\re-ranking};
\node[process, below right=0.1cm and 0.75cm of rerank] (diverse) {Filters, deduplication\\and company diversity};
\node[output, right=0.9cm of diverse] (explain) {Explainable ranked\\job recommendations};

\draw[arrow] (query) -- (qproc);
\draw[arrow] (qproc) -- (qenc);
\draw[arrow] (qenc) -- (retrieval);
\draw[arrow] (retrieval) -- (ranking);
\draw[arrow] (ranking) -- (rerank);
\draw[arrow] (rerank) -- node[above, font=\tiny]{yes} (ce);
\draw[arrow] (ce) |- (diverse);
\draw[arrow] (rerank) -- node[below, font=\tiny]{no} (diverse);
\draw[arrow] (diverse) -- (explain);

\draw[dashedarrow] (sparse.south) |- (ranking.north);
\draw[dashedarrow] (dense.south) |- (retrieval.north);

\node[group, fit=(raw)(prep)(doc)(tfidf)(sbert)(sparse)(dense), label={[font=\scriptsize\bfseries]above:Offline Index Construction}] {};
\node[group, fit=(query)(qproc)(qenc)(retrieval)(ranking)(rerank)(ce)(diverse)(explain), label={[font=\scriptsize\bfseries]below:Online Recommendation Flow}] {};
\end{tikzpicture}%
}
\caption{Overall architecture of the proposed intelligent job recommendation system.}
\label{fig:architecture}
\end{figure*}
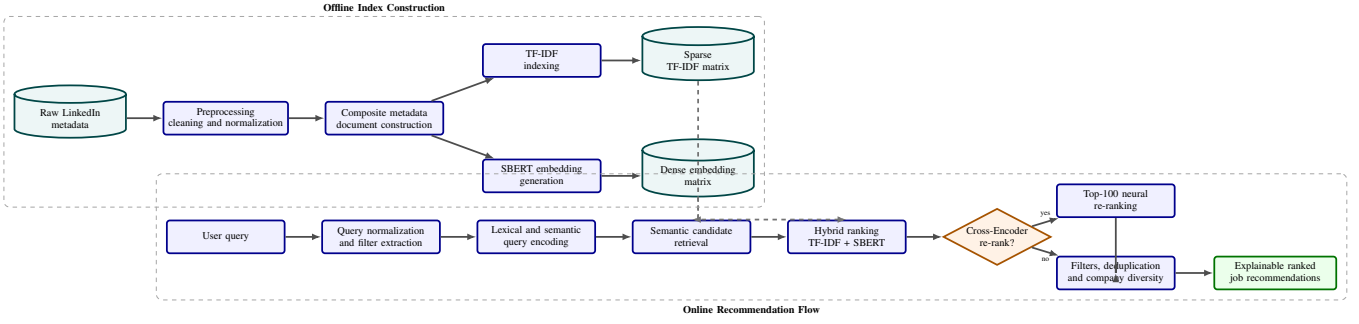

\section{Retrieval Pipeline}

The retrieval pipeline combines semantic candidate generation with hybrid ranking. Semantic retrieval is applied first because it can identify related postings even when exact lexical overlap is limited. Lexical evidence is then reintroduced during ranking to preserve exact matching for important query constraints.

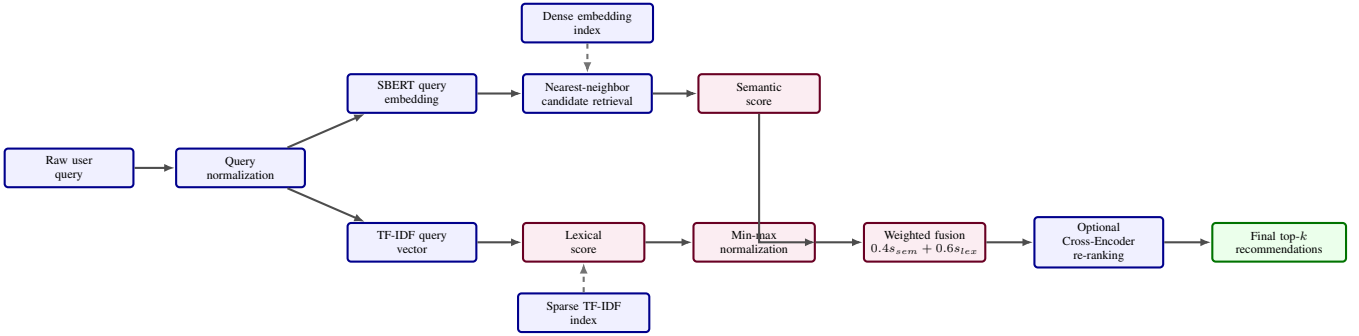
\begin{figure*}[t]
\centering
\resizebox{0.98\textwidth}{!}{%
\begin{tikzpicture}[
    node distance=0.75cm and 0.82cm,
    stage/.style={rectangle, rounded corners=2pt, draw=blue!55!black, fill=blue!6, very thick, align=center, minimum width=2.6cm, minimum height=0.78cm, font=\scriptsize},
    score/.style={rectangle, rounded corners=2pt, draw=purple!55!black, fill=purple!7, very thick, align=center, minimum width=2.45cm, minimum height=0.78cm, font=\scriptsize},
    output/.style={rectangle, rounded corners=2pt, draw=green!45!black, fill=green!8, very thick, align=center, minimum width=2.7cm, minimum height=0.78cm, font=\scriptsize},
    arrow/.style={-{Latex[length=2.2mm]}, very thick, draw=black!70},
    dashedarrow/.style={-{Latex[length=2.2mm]}, very thick, dashed, draw=black!55}
]
\node[stage] (query) {Raw user\\query};
\node[stage, right=of query] (norm) {Query\\normalization};
\node[stage, above right=0.68cm and 0.82cm of norm] (sbertq) {SBERT query\\embedding};
\node[stage, below right=0.68cm and 0.82cm of norm] (tfidfq) {TF-IDF query\\vector};
\node[stage, right=0.9cm of sbertq] (nn) {Nearest-neighbor\\candidate retrieval};
\node[score, right=0.9cm of nn] (semscore) {Semantic\\score};
\node[score, right=0.9cm of tfidfq] (lexscore) {Lexical\\score};
\node[score, right=0.95cm of lexscore] (normscores) {Min-max\\normalization};
\node[score, right=0.95cm of normscores] (fusion) {Weighted fusion\\$0.4s_{sem}+0.6s_{lex}$};
\node[stage, right=0.95cm of fusion] (rerank) {Optional\\Cross-Encoder\\re-ranking};
\node[output, right=0.95cm of rerank] (topk) {Final top-$k$\\recommendations};

\draw[arrow] (query) -- (norm);
\draw[arrow] (norm) -- (sbertq);
\draw[arrow] (norm) -- (tfidfq);
\draw[arrow] (sbertq) -- (nn);
\draw[arrow] (nn) -- (semscore);
\draw[arrow] (tfidfq) -- (lexscore);
\draw[arrow] (semscore) |- (normscores);
\draw[arrow] (lexscore) -- (normscores);
\draw[arrow] (normscores) -- (fusion);
\draw[arrow] (fusion) -- (rerank);
\draw[arrow] (rerank) -- (topk);

\node[stage, above=0.6cm of nn, minimum width=2.65cm] (denseidx) {Dense embedding\\index};
\node[stage, below=0.6cm of lexscore, minimum width=2.65cm] (sparseidx) {Sparse TF-IDF\\index};
\draw[dashedarrow] (denseidx) -- (nn);
\draw[dashedarrow] (sparseidx) -- (lexscore);
\end{tikzpicture}%
}
\caption{Hybrid retrieval pipeline combining semantic candidate generation and lexical-semantic ranking.}
\label{fig:retrieval_pipeline}
\end{figure*}

\subsection{Query-Aware Filtering and Diversification}

The system extracts explicit filters from the query when possible. Supported filters include employment type, seniority level, location hints, and work mode. If the extracted filters remove all candidates, the system falls back to the unfiltered candidate set to avoid returning an empty result list.

The final ranking stage also applies duplicate suppression using a tuple of normalized job title, company, and location. A company-level cap limits repeated results from the same employer. These rules improve result diversity without changing the core retrieval model.

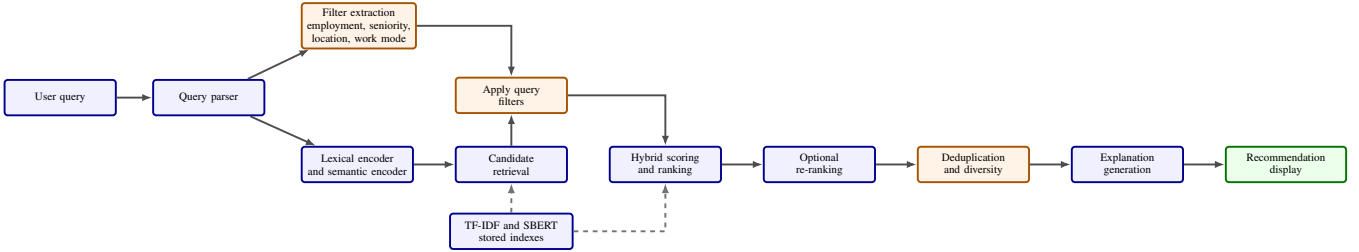
\begin{figure*}[t]
\centering
\resizebox{0.98\textwidth}{!}{%
\begin{tikzpicture}[
    node distance=0.72cm and 0.78cm,
    box/.style={rectangle, rounded corners=2pt, draw=blue!55!black, fill=blue!6, very thick, align=center, minimum width=2.45cm, minimum height=0.78cm, font=\scriptsize},
    filter/.style={rectangle, rounded corners=2pt, draw=orange!65!black, fill=orange!10, very thick, align=center, minimum width=2.45cm, minimum height=0.78cm, font=\scriptsize},
    output/.style={rectangle, rounded corners=2pt, draw=green!45!black, fill=green!8, very thick, align=center, minimum width=2.65cm, minimum height=0.78cm, font=\scriptsize},
    arrow/.style={-{Latex[length=2.2mm]}, very thick, draw=black!70},
    dashedarrow/.style={-{Latex[length=2.2mm]}, very thick, dashed, draw=black!55}
]
\node[box] (user) {User query};
\node[box, right=of user] (parser) {Query parser};
\node[filter, above right=0.65cm and 0.78cm of parser] (filters) {Filter extraction\\employment, seniority,\\location, work mode};
\node[box, below right=0.65cm and 0.78cm of parser] (encoders) {Lexical encoder\\and semantic encoder};
\node[box, right=0.9cm of encoders] (candidates) {Candidate\\retrieval};
\node[filter, above=0.7cm of candidates] (applyfilters) {Apply query\\filters};
\node[box, right=0.9cm of candidates] (scoring) {Hybrid scoring\\and ranking};
\node[box, right=0.9cm of scoring] (rerank) {Optional\\re-ranking};
\node[filter, right=0.9cm of rerank] (diversity) {Deduplication\\and diversity};
\node[box, right=0.9cm of diversity] (explain) {Explanation\\generation};
\node[output, right=0.9cm of explain] (display) {Recommendation\\display};

\draw[arrow] (user) -- (parser);
\draw[arrow] (parser) -- (filters);
\draw[arrow] (parser) -- (encoders);
\draw[arrow] (encoders) -- (candidates);
\draw[arrow] (filters) -| (applyfilters);
\draw[arrow] (candidates) -- (applyfilters);
\draw[arrow] (applyfilters) -| (scoring);
\draw[arrow] (scoring) -- (rerank);
\draw[arrow] (rerank) -- (diversity);
\draw[arrow] (diversity) -- (explain);
\draw[arrow] (explain) -- (display);

\node[box, below=0.65cm of candidates, minimum width=2.7cm] (indexes) {TF-IDF and SBERT\\stored indexes};
\draw[dashedarrow] (indexes) -- (candidates);
\draw[dashedarrow] (indexes) -| (scoring);
\end{tikzpicture}%
}
\caption{Online query flow from user input to explainable job recommendations.}
\label{fig:query_flow}
\end{figure*}

\section{Explainable AI Layer}

The explainability component is designed to make recommendations understandable to end users. Instead of exposing only a numerical score, the system reports evidence that links the user query to the recommended posting. This form of explanation is consistent with feature-based explanation strategies for recommender systems \cite{zhang2020explainable}.

The explanation layer includes four forms of evidence:
\begin{itemize}
    \item \textit{Keyword overlap}: shared terms between the user query and the composite job document.
    \item \textit{Filter explanation}: extracted constraints such as remote, hybrid, junior, full-time, or a location term.
    \item \textit{Metadata evidence}: matches in job function, industry, seniority level, or employment type.
    \item \textit{Ranking evidence}: indication of whether a result was mainly supported by lexical matching, semantic similarity, or neural re-ranking.
\end{itemize}

For example, for the query ``remote junior data analyst London'', a recommended job may include an explanation showing matched keywords such as ``data'', ``analyst'', ``remote'', and ``London'', together with applied filters for work mode, seniority, and location. This explanation does not fully interpret the internal embedding space, but it provides practical transparency based on metadata visible to users.

\section{Dataset and Preprocessing}

The evaluation used a LinkedIn job posting dataset. The raw dataset contained 31,597 records. After cleaning and validation, 31,262 valid records remained. The system used only structured metadata fields: job title, company name, location, hiring status, date, seniority level, job function, employment type, and industry.

The preprocessing pipeline included missing-value handling, whitespace normalization, job-title normalization, abbreviation expansion, and removal of noisy records. Examples of abbreviation expansion include ML to machine learning, AI to artificial intelligence, SWE and SDE to software engineer, QA to quality assurance, PM to product manager, and MLE to machine learning engineer.

\begin{table}[t]
\centering
\caption{Dataset fields used by the metadata-only recommendation system.}
\label{tab:dataset_fields}
\footnotesize
\begin{tabular}{@{}p{0.28\linewidth}p{0.58\linewidth}@{}}
\toprule
\textbf{Field} & \textbf{Role in the system} \\
\midrule
Job title & Primary role signal and strongest lexical evidence \\
Company name & Employer metadata and duplicate-control feature \\
Location & Geographic and work-location matching \\
Hiring status & Structured posting status metadata \\
Date & Temporal metadata retained from the source dataset \\
Seniority level & Career-level filtering and relevance grading \\
Job function & Functional category for matching and evaluation \\
Employment type & Full-time, part-time, contract, internship, and related filters \\
Industry & Domain-level evidence for matching and relevance grading \\
\bottomrule
\end{tabular}
\end{table}

\begin{table}[t]
\centering
\caption{Dataset size before and after preprocessing.}
\label{tab:dataset_size}
\footnotesize
\begin{tabular}{@{}lcl@{}}
\toprule
\textbf{Stage} & \textbf{Records} & \textbf{Description} \\
\midrule
Raw dataset & 31,597 & Original LinkedIn job postings \\
Cleaned dataset & 31,262 & Valid records after preprocessing \\
Removed records & 335 & Invalid, noisy, or unusable records \\
\bottomrule
\end{tabular}
\end{table}

\section{Experimental Evaluation}

The evaluation used an internal metadata-based relevance protocol because human relevance judgments and user interaction logs were unavailable. Seed jobs were sampled from the dataset, and their normalized job titles were used as queries. Retrieved jobs were compared with the seed job using metadata consistency.

Relevance labels were assigned as follows: 3 for the same normalized job title, 2 for the same job function or industry, 1 for the same seniority level or employment type, and 0 otherwise. Precision@10 considered labels 2 and 3 as relevant. nDCG@10 was used to evaluate ranking quality with graded relevance.

Precision@10 is defined as:
\begin{equation}
\mathrm{Precision@10}=\frac{|\{d_i \in R_{10}: rel(d_i)\geq 2\}|}{10},
\end{equation}
where $R_{10}$ is the set of the top 10 returned jobs.

Discounted cumulative gain at rank $k$ is computed as:
\begin{equation}
\mathrm{DCG@}k=\sum_{i=1}^{k}\frac{2^{rel_i}-1}{\log_2(i+1)}.
\end{equation}
The normalized version is:
\begin{equation}
\mathrm{nDCG@}k=\frac{\mathrm{DCG@}k}{\mathrm{IDCG@}k},
\end{equation}
where $\mathrm{IDCG@}k$ is the ideal DCG obtained by sorting results by true graded relevance.

\begin{table}[t]
\centering
\caption{Internal relevance grading protocol used for offline evaluation.}
\label{tab:relevance}
\footnotesize
\begin{tabular}{@{}cp{0.49\linewidth}p{0.27\linewidth}@{}}
\toprule
\textbf{Grade} & \textbf{Condition} & \textbf{Interpretation} \\
\midrule
3 & Same normalized job title & Highly relevant \\
2 & Same job function or same industry & Relevant \\
1 & Same seniority level or same employment type & Weakly related \\
0 & None of the above metadata matches & Not relevant \\
\bottomrule
\end{tabular}
\end{table}

\subsection{Hybrid Retrieval Results}

Table~\ref{tab:hybrid_results} reports the hybrid retrieval results across different candidate sizes and weighting settings. The best reported setting used 250 semantic candidates with semantic weight 0.4 and lexical weight 0.6, achieving Precision@10 of 0.8032 and nDCG@10 of 0.9496.

\begin{table}[t]
\centering
\caption{Hybrid retrieval performance under different candidate sizes and weighting configurations.}
\label{tab:hybrid_results}
\footnotesize
\begin{tabular}{@{}ccccc@{}}
\toprule
\textbf{Candidates} & \textbf{Sem.} & \textbf{Lex.} & \textbf{P@10} & \textbf{nDCG@10} \\
\midrule
80  & 0.7 & 0.3 & 0.7624 & 0.9360 \\
80  & 0.6 & 0.4 & 0.7676 & 0.9388 \\
80  & 0.5 & 0.5 & 0.7780 & 0.9392 \\
80  & 0.4 & 0.6 & 0.7884 & 0.9432 \\
150 & 0.7 & 0.3 & 0.7640 & 0.9365 \\
150 & 0.6 & 0.4 & 0.7776 & 0.9413 \\
150 & 0.5 & 0.5 & 0.7844 & 0.9420 \\
150 & 0.4 & 0.6 & 0.7984 & 0.9478 \\
250 & 0.7 & 0.3 & 0.7680 & 0.9393 \\
250 & 0.6 & 0.4 & 0.7832 & 0.9389 \\
250 & 0.5 & 0.5 & 0.7936 & 0.9454 \\
250 & 0.4 & 0.6 & \textbf{0.8032} & \textbf{0.9496} \\
\bottomrule
\end{tabular}
\end{table}

\subsection{Cross-Encoder Re-ranking Results}

Table~\ref{tab:rerank_results} compares the baseline hybrid ranking with the optional Cross-Encoder re-ranking stage. Re-ranking was applied to the top 100 candidates using $\alpha=0.7$. The Cross-Encoder produced modest improvements in both metrics under the internal evaluation protocol.

\begin{table}[t]
\centering
\caption{Baseline hybrid ranking versus Cross-Encoder re-ranking.}
\label{tab:rerank_results}
\footnotesize
\begin{tabular}{@{}lccc@{}}
\toprule
\textbf{Configuration} & \textbf{P@10} & \textbf{nDCG@10} & \textbf{Notes} \\
\midrule
Baseline hybrid & $0.7896 \pm 0.2896$ & $0.9666 \pm 0.1051$ & No Cross-Encoder \\
Cross-Encoder & $0.7948 \pm 0.2946$ & $0.9739 \pm 0.1046$ & Top 100 candidates \\
\midrule
Delta & +0.0052 & +0.0072 & Rerank minus baseline \\
\bottomrule
\end{tabular}
\end{table}

\section{Discussion}

The results show that a hybrid retrieval design is suitable for metadata-only job recommendation under the evaluation protocol used in this study. Larger semantic candidate sets improved the likelihood that relevant postings were available to the ranking stage. At the same time, configurations with stronger lexical weighting performed best. This result is consistent with the short and structured nature of the metadata, where exact terms in job titles, locations, seniority levels, and employment types remain informative.

The Cross-Encoder re-ranking stage improved both Precision@10 and nDCG@10, although the gains were moderate. This outcome is expected because the baseline hybrid model already performed strongly under the metadata-derived relevance protocol. The benefit of re-ranking is its more detailed query-candidate interaction, while its drawback is increased computational cost \cite{lin2021pretrained}.

The explanation layer provides practical transparency by showing matched keywords and applied filters. The approach does not claim to fully explain dense embedding behavior. Instead, it exposes metadata-based evidence that is directly connected to visible job attributes and user query terms.

\subsection{Limitations}

This study has several limitations. First, the system uses structured metadata only and does not include full job descriptions, skill requirements, salary, education level, or company descriptions. Second, the evaluation uses heuristic relevance labels derived from metadata consistency rather than human judgments or real user interactions. Third, the models are pre-trained general-purpose models and were not fine-tuned on a job-specific relevance dataset. Fourth, Cross-Encoder re-ranking improves quality but increases latency and computational cost. Finally, the dataset represents a fixed snapshot of job postings, while operational recruitment platforms require continuous updates as jobs are posted, modified, and closed.

\section{Conclusion}

This paper presented an intelligent job recommendation system that combines semantic retrieval, lexical matching, explainable AI techniques, and optional neural re-ranking. The system was designed for a metadata-only setting where full descriptions and user interaction histories are unavailable. The proposed pipeline uses TF-IDF for exact lexical evidence, Sentence-BERT for semantic candidate retrieval, weighted hybrid scoring for ranking, query-aware filters for explicit constraints, and explanation fields for transparency.

Experiments on a cleaned LinkedIn job dataset showed that the best hybrid configuration achieved Precision@10 of 0.8032 and nDCG@10 of 0.9496. Cross-Encoder re-ranking further improved Precision@10 from 0.7896 to 0.7948 and nDCG@10 from 0.9666 to 0.9739 under the internal evaluation protocol. These findings support the conclusion that effective and interpretable metadata-only job recommendation is feasible when preprocessing, retrieval, ranking, filtering, and evaluation are carefully controlled.

Future work should include human relevance judgments, real user interaction data, full job descriptions, skill extraction, learning-to-rank models \cite{burges2005ranking}, fairness analysis, and deployment-oriented latency evaluation.

\bibliographystyle{IEEEtran}
\bibliography{references}

\end{document}